\documentclass[lettersize,journal]{IEEEtran}

\usepackage{amsmath,amsfonts}
\usepackage{algorithmic}
\usepackage{algorithm}
\usepackage{array}
\usepackage[caption=false,font=footnotesize,farskip=0pt]{subfig}
\usepackage{textcomp}
\usepackage{stfloats}
\usepackage{url}
\usepackage{verbatim}
\usepackage{graphicx}
\usepackage{cite}

\usepackage[english]{babel}
\usepackage[T1]{fontenc}
\usepackage[cmintegrals]{newtxmath}
\usepackage{bm}
\usepackage{multirow}
\usepackage{enumitem}
\usepackage{soul}
\usepackage{booktabs}

\usepackage[acronym]{glossaries}
\newacronym{3gpp}{3GPP}{3rd Generation Partnership Project}
\newacronym{5g}{5G}{the 5th generation of mobile networks}
\newacronym{6g}{6G}{sixth generation of mobile networks}
\newacronym{awgn}{AWGN}{additive white Gaussian noise}
\newacronym{b5g}{B5G}{5G and Beyond}
\newacronym{ber}{BER}{bit error rate}
\newacronym{bler}{BLER}{block error rate}
\newacronym{bs}{BS}{base station}
\newacronym{cdf}{CDF}{cumulative distribution function}
\newacronym{cml}{CML}{commercial microwave link}
\newacronym{crb}{CRB}{Cramér-Rao bound}
\newacronym{csi}{CSI}{channel state information}
\newacronym{dl}{DL}{downlink}
\newacronym{dmrab}{DMRAB}{disjoint matching and resource allocation benchmark}
\newacronym{embb}{eMBB}{Enhanced Mobile Broadband}
\newacronym{gap}{GAP}{Generalized Assignment Problem}
\newacronym{geo}{GEO}{Geosynchronous Earth Orbit}
\newacronym{gsl}{GSL}{Ground-to-Satellite Link}
\newacronym{iot}{IoT}{Internet of Things}
\newacronym{isac}{ISAC}{integrated sensing and communications}
\newacronym{isl}{ISL}{Inter-Satellite Link}
\newacronym{jmra}{JMRA}{joint matching and resource allocation}
\newacronym{kpi}{KPI}{key performance indicator}
\newacronym{leo}{LEO}{Low Earth Orbit}
\newacronym{mcs}{MCS}{modulation and coding scheme}
\newacronym{mgap}{MGAP}{Multi-Level Generalized Assignment Problem}
\newacronym{milp}{MILP}{mixed-integer linear problem}
\newacronym{mimo}{MIMO}{Multiple-Input Multiple-Output}
\newacronym{ml}{ML}{machine learning}
\newacronym{mle}{MLE}{maximum likelihood estimator}
\newacronym{mmtc}{mMTC}{massive-Machine Type Communications}
\newacronym{mse}{MSE}{mean-squared error}
\newacronym{nmse}{NMSE}{normalized mean-squared error}
\newacronym{noma}{NOMA}{non-orthogonal multiple access}
\newacronym{nr}{NR}{New Radio}
\newacronym{ntn}{NTN}{Non-Terrestrial Network}
\newacronym{ofdma}{OFDMA}{Orthogonal Frequency-Division Multiple Access}
\newacronym{pdf}{PDF}{probability density function}
\newacronym{pdv}{PDV}{packet delay variation}
\newacronym{pmf}{PMF}{probability mass function}
\newacronym{ppp}{PPP}{Poisson point process}
\newacronym{psv}{PSV}{probability of simultaneity violation}
\newacronym{qos}{QoS}{Quality of Service}
\newacronym{ra}{RA}{resource allocation}
\newacronym{ran}{RAN}{radio access network}
\newacronym{rssi}{RSSI}{received signal strength indicator}
\newacronym{rtt}{RTT}{round-trip time}
\newacronym{rv}{RV}{random variable}
\newacronym{snr}{SNR}{signal-to-noise ratio}
\newacronym{sr}{SR}{scheduling request}
\newacronym{ss}{SS}{synchronization signal}
\newacronym{tdd}{TDD}{time-division duplexing}
\newacronym{twi}{TWI}{temporal window of integration}
\newacronym{uav}{UAV}{unmanned aerial vehicle}
\newacronym{ue}{UE}{User Equipment}
\newacronym{ul}{UL}{uplink}

\newtheorem{lemma}{Lemma}
\newtheorem{definition}{Definition}

\usepackage{stfloats}
\fnbelowfloat

\begin{document}

\title{Preserving Simultaneity and Chronology for Sensing in Perceptive Wireless Networks}

\author{João Henrique Inacio de Souza,~\IEEEmembership{Member,~IEEE,} Fabio Saggese,~\IEEEmembership{Member,~IEEE,}\\ Beatriz Soret,~\IEEEmembership{Senior Member,~IEEE,} and Petar Popovski,~\IEEEmembership{Fellow,~IEEE}
\thanks{J. H. Inacio de Souza, F. Saggese, B. Soret, and P. Popovski are with the Department of Electronic Systems, Aalborg University, Denmark (e-mail: \{jhids,fasa,bsa,petarp\}@es.aau.dk). B. Soret is also with the Telecommunications Research Institute, Universidad de Málaga, Spain. This work was supported in part by the Villum Investigator Grant “WATER” from the Velux Foundation, Denmark, and in part by the SNS JU project 6G-GOALS under the EU’s Horizon Europe program under Grant Agreement No 101139232.}}

\maketitle

\begin{abstract}
    We address the challenge of preserving the simultaneity and chronology of sensing events in multisensor systems with wireless links. The network uses temporal windows of integration (TWIs), borrowed from human multisensory perception, to preserve the temporal structure of the sensing data at the application side. We introduce a composite latency model for propagation, sensing, and communication that leads to the derivation of the probability of simultaneity violation. This is used to select the TWI duration aiming to achieve the desired degrees of chronological preservation, while maintaining the throughput of events. The letter provides important insights and analytical tools about the TWI impact on the event registration.
\end{abstract}

\begin{IEEEkeywords}
    Internet of Things (IoT), logical clocks, perceptive networks.
\end{IEEEkeywords}

%
%
\section{Introduction}\label{sec:introduction}

\IEEEPARstart{T}{HE} evolution towards perceptive mobile networks~\cite{Xie2023}, together with edge-computing and AI-native technologies, is creating a platform of unprecedented tools to monitor and control cyber-physical systems. Going beyond radio sensing, \glspl{bs} are expected to get multimodal inputs from sensors and \gls{iot} devices, supporting complex learning/inference tasks and capturing associations between data modalities for insights into physical processes~\cite{Baltrusaitis2019}.
Fusing distributed multimodal sensor data enables \emph{perceptive wireless networks} capable of interpreting the physical environment in ways analogous to human perception.
Multimodal processing has been shown to improve the performance of, \emph{e.g.}, target tracking systems~\cite{Hao2023}.
However, multimodal sensor fusion is prone to the reception of \emph{out-of-order measurements} due to random delays induced by, \emph{e.g.}, communication. In estimation and inference tasks, outdated measurements received out-of-order can be leveraged by applying \emph{retrodiction} followed by an update step~\cite{GarciaFernandez2021}. However, such a strategy is not suitable for decision-making, as different execution paths may be taken depending on the observations. Resilient task completion requires preservation of event chronology to process all the measurements in their original order.

In distributed systems, message ordering protocols based on \emph{logical clocks} can preserve the correct chronology of sensor updates, ensuring delivery to the application in their original order~\cite{vanSteen2024}. Nonetheless, such protocols introduce significant overhead for sharing vector clocks, while they cannot handle physical events before they are converted to the digital domain by, \emph{e.g.}, sensing~\cite{Popovski2024}. Alternatively, when the delay to receive a sensor update has a known upper bound~\cite{Romer2003}, it can be used to set a deadline to start processing only after receiving all updates. Similarly, but based on mechanisms that enable the human multisensory perception, the \gls{twi} framework~\cite{Popovski2024} proposes the establishment of timing constraints to process events, guaranteeing their correct temporal ordering and tracking the possible causal relations between them. A related work~\cite{Chen2024} considers out-of-order arrivals from the Age of Information perspective.

Inspired by the \gls{twi} framework, we address the problem of event chronology when receiving status updates from wireless sensors that observe the same physical process, delivering to an application hosted at a \gls{bs}. Our main contribution is the development of analytical tools that link delays to the event chronology as perceived by the application. We extend~\cite{Popovski2024} by developing a model that covers the propagation of physical signals generated by the monitored process, the computation delay involved in the sensing task, and the communication delay to convey the updates to the \gls{bs}. By applying this model, we derive the \gls{psv} in a scenario where event-driven updates related to a physical process \emph{must be delivered concurrently to the application}. Then, we design the \gls{twi} duration to provide statistical guarantees that the delivery to the application of status updates generated by the same physical event will attain a target \gls{psv}. The analysis is supplemented by numerical results that reveal trends in how \gls{twi} choice impacts simultaneity preservation.

%
%

\begin{figure}[t]
    \centering
    \includegraphics[width=0.9\columnwidth]{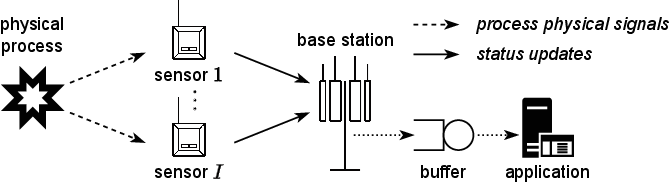}
    \caption{System to monitor a physical process from event-driven status updates produced by wireless sensors}
    \label{fig:scenario}
\end{figure}

\section{System Model}\label{sec:system-model}

\begin{figure*}[t]
    \centering
    \includegraphics[width=.9\textwidth]{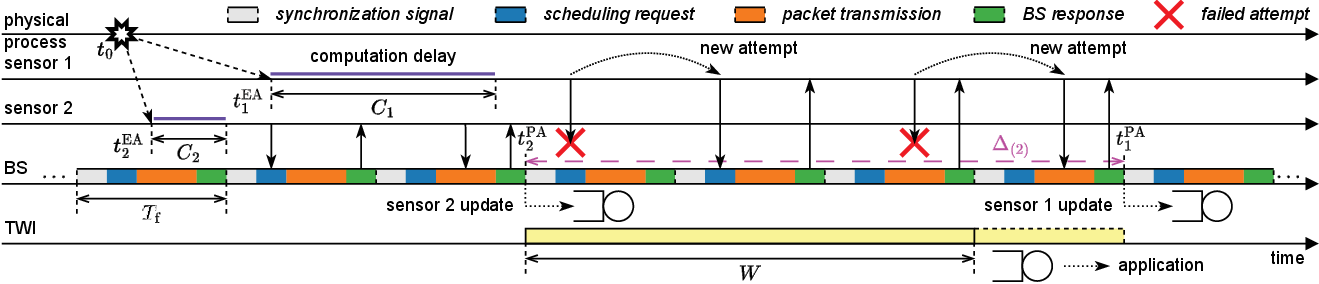}
    \caption{Timing diagram illustrating a physical process event, the generation and transmission of status updates related to that event by two sensors, and the delivery of the updates to the application based on the \gls{twi} framework. In this scenario, sensor~1's update arrived later than sensor~2's due to the longer propagation and computation delays, and the delay resulting from the failed \acrshort{sr} and packet transmission attempt. In this example, the \gls{twi} design is inadequate for the simultaneous delivery of the updates: the \gls{twi} lasts for three frames, while the update arrival times are separated by four frames.}
    \label{fig:timing-diagram}
\end{figure*}

We consider the scenario in Fig.~\ref{fig:scenario}, in which an application hosted at the \gls{bs} monitors a physical process based on event-driven status updates sent by $I\in\mathbb{N}_+$ wireless sensors. When the physical process generates an event, the sensors produce and transmit updates regarding it. All the updates resulting from the same physical event need to be considered simultaneous from the application perspective. Therefore, the \gls{bs} must ensure that all received updates will be delivered concurrently to the application. To do so, the \gls{bs} stores the updates in a buffer, delivering them and cleaning the buffer under two situations: 1) when all $I$ updates have been successfully received and stored, or 2) when $W>0$ seconds have passed after the reception and storage of the first update. Hence, $W$ denotes the maximum time window at which the application will perceive status updates as simultaneous and, accordingly, is the duration of the application \gls{twi}~\cite{Popovski2024}.

\begin{definition}[Simultaneity violation] \label{def:sv}
    A simultaneity violation event occurs at the application when updates related to the same physical event are delivered in different \glspl{twi}. We refer to its probability as the \gls{psv}.
\end{definition}

To design the \gls{twi} duration $W$, we aim to link it to the \gls{psv}; this is done by evaluating the \gls{pdv}.

\begin{definition}[Packet delay variation] \label{def:pdv}
    The \gls{pdv} is defined as the time interval between the reception at the \gls{bs} of the first (fastest) and last (slowest) status updates.
\end{definition}

The investigation is carried out under the sensing and communication processes depicted in Fig.~\ref{fig:timing-diagram} and detailed below.

%
%

\subsubsection{Sensing Process} The monitored process generates events in the form of physical signals (\emph{e.g.}, radio, acoustic, seismic, optical), propagating toward sensors at the speed $v>0$.
For simplicity, only one type of physical signal is considered.
We assume the time interval between two consecutive events to be much larger than the application's \gls{twi}, such that new events do not occur while the \gls{bs} receives status updates produced by an event. The time an event occurred is denoted by $t_0 \geq 0$. Sensors are uniformly distributed over a circular region with radius $D_{\max}>0$, centered at the position of the monitored process. The \gls{pdf} of the distance from the $i-$th sensor to the physical process is
\begin{equation}
    \label{eq:pdf-D}
    f_D(d) = \frac{2d}{D_{\max}^2}, \,\, d \in [0, D_{\max}].
\end{equation}
Since the physical signals generated by the process events travel at a limited speed, depending on the distance $D_i$, the propagation delay cannot be neglected. Accordingly, we define the \emph{event action time} of sensor $i$ as the time at which the physical signals arrive such a sensor, \emph{i.e.},
\begin{equation}
    t_i^\text{EA}=t_0 + \frac{D_i}{v}.
\end{equation}

The sensors produce event-driven status updates based on observations of the physical signals, following a change-aware sampling and communication policy~\cite{Pappas2021}. A status update comprises a data packet containing $b\in\mathbb{N}_+$ bits. To produce a single update, a sensor induces a computation delay $C_i \sim \mathcal{U}(C_{\min}, C_{\max})$, where $0<C_{\min}<C_{\max}$ are the best- and worst-case computation delays, respectively. This stochastic model reflects the device heterogeneity in larger sensor populations, where irregular computation delays can be due to, \emph{e.g.}, different hardware platforms.

%
%

\subsubsection{Communication Process} The sensors send the status updates through wireless transmissions dynamically scheduled by the \gls{bs}. We consider a narrowband frame-based communication process of bandwidth $B > 0$ and frame duration $T_\text{f}>0$ as in Fig.~\ref{fig:timing-diagram}. Each frame has \gls{dl} and \gls{ul} sub-frames: \gls{ss} (\gls{dl}), \gls{sr} (\gls{ul}), packet transmission (\gls{ul}), and \gls{bs} response (\gls{dl}). We assume a block-fading channel with channel coefficient $h_i \sim \mathcal{CN}(0, \beta_i)$ between the \gls{bs} and sensor $i$, where $\beta_i$ is the large-scale fading, known for all $I$ sensors.

Transmissions are scheduled upon random access requests made individually by the sensors after producing their status updates. To send a request, sensor $i$ first listens to a \gls{ss} to acquire frame synchronization; then, during the \gls{sr} subframe, it transmits the preamble $\bm{s}_i \in \mathbb{C}^S$ such that $S\geq I$ and $\|\bm{s}_i\|=1$, with transmit power $P_i>0$. Such a preamble defines a unique identifier for each sensor and is taken from a set of mutually orthogonal sequences, \emph{i.e.}, $\bm{s}_i^\htransp\bm{s}_j = 0,~\forall i\neq j$. Therefore, no preamble collision occurs if multiple sensors send simultaneous \glspl{sr}. Let $\mathcal{I}$ denote the indices of the sensors that sent an \gls{sr} during a specific frame and $\bm{n}\sim\mathcal{CN}(\bm{0}, N_0B\bm{I})$ the thermal noise at the \gls{bs} receiver, with $N_0>0$ being the noise spectral density. The \gls{ul} received signal at the \gls{bs} containing the \gls{sr} preambles is denoted as 
$\bm{y} = \sum_{i \in \mathcal{I}} h_i \sqrt{P_i} \bm{s}_i + \bm{n}$.
While no collision can happen in our setting, the \gls{bs} needs to detect transmitted scheduling preambles to decode them. We assume the \gls{bs} employs a maximum likelihood detection process on the normalized received signal correlated with the preamble, \emph{i.e.}, $\hat{y}_i = \bm{s}_i^\htransp\bm{y}/\sqrt{P_iN_0B}$. Using standard detection theory~\cite{Kay1997detection}, this processing yields the optimal threshold $\eta_i = (1+\gamma_i^{-1}) \ln(1+\gamma_i)$, where $\gamma_i = P_i\beta_i/(N_0B)$ is the average received \gls{snr} at the \gls{bs}.
When the \gls{bs} detects the preambles transmitted by a generic sensor~$i$, \emph{i.e.}, whenever $\hat{y}_i \ge \eta_i$, for any $i \in\{1, \dots, I\}$, it sends a scheduling grant in the \gls{bs} response sub-frame intended for the preamble's owner. The scheduling grant is assumed orthogonal among users. A scheduling grant allows a sensor to transmit its status update in the next communication frame, within the packet transmission sub-frame. If a sensor that sent an \gls{sr} does not find a grant intended for it in the \gls{bs} response, it will make a new attempt in the subsequent frame. Each sensor can send an \gls{sr} up to $M_{\max}\in\mathbb{N}_+$ times, including the first attempt. The number of attempts sent by sensor $i$ is denoted by $m_i\leq M_{\max}$ and a packet is dropped when the sensor does not receive a scheduling grant after the $M_{\max}$-th attempt.

After receiving a scheduling grant in a frame, the sensor uses the packet transmission opportunity of the next frame to send its status update. If more than one sensor got a scheduling grant in a frame, then the \gls{bs} uses round-robin scheduling to accommodate all packet transmissions. Let $0<T_\text{p}<T_\text{f}$ denote the time interval of the packet transmission opportunity in a frame. We have assumed a quasi-static fading channel which has zero dispersion~\cite{Yang2014quasistatic}; thus, finite blocklength effects are negligible and the outage probability accurately captures the error performance.
The transmission of sensor $i$ is in outage if the instantaneous \gls{snr} measured at the \gls{bs} is less than $\gamma_\text{TH} = 2^{b/(T_\text{p}B)} - 1$. Therefore, the outage probability is: 
\begin{equation}
    \label{eq:outage-probability}
    \epsilon_i = \Pr\left\{ \frac{P_i|h_i|^2}{N_0B} < \gamma_\text{TH} \right\}
    =1 - \exp\left\{- \frac{\gamma_\text{TH}}{\gamma_i}\right\},
\end{equation}
given that $|h_i|^2\sim\mathrm{Exp}(1/\beta_i)$.
When a transmission does not incur an outage, the \gls{bs} successfully receives the packet and sends an acknowledgment to the sensor in the \gls{bs} response of the same frame. Otherwise, if an outage occurs, the \gls{bs} provide a new scheduling grant for the transmitting sensor in the \gls{bs} response sub-frame; the sensor can attempt the packet transmission during the subsequent frame. A scheduled sensor can make up to $N_{\max}\in\mathbb{N}_+$ transmission attempts before dropping its packet, while we denote the number of transmission attempts made by sensor $i$ by $n_i\leq N_{\max}$. $N_{\max}$ is not necessarily equal to $M_{\max}$.

From the system model, one can perceive that the propagation of the physical signals generated by the monitored process, the computation by the sensors, and the communication with the \gls{bs} introduce random delays in the reception of status updates produced by an event. To design the application \gls{twi} duration, we need to understand how these delays affect the arrival times of the first and last received status updates at the \gls{bs}, as investigated in the following section.

%
%
\section{TWI Design Based on the Distribution of the Packet Delay Variation}\label{sec:twi-design}

In this section, we first develop statistical models for the sensors' packet arrival time at the \gls{bs} and the \gls{pdv} between the first and last received status updates. Then, we design the application \gls{twi} duration based on the characteristics of the physical, sensing, and communication processes.

We analyzed three timing components: 1)~the physical event action time, 2)~the sensor computation delay, and 3)~the communication delay. All results are conditioned to the event time $t_0$ and propagation velocity $v$, the average \glspl{snr} $\{\gamma_i\}_{i=1}^I$, and the communication parameters $T_\text{f}$, $M_{\max}$, and $N_{\max}$.

When conditioned to $t_0$ and $v$, the \gls{pdf} of $t_i^\text{EA}$ depends only on the \gls{pdf} of $D_i$. Hence, from eq.~\eqref{eq:pdf-D}, the \gls{pdf} of $t_i^\text{EA}$ is
\begin{equation}
    \label{eq:distribution-action-time}
    t_i^\text{EA} \sim \frac{2v^2(t-t_0)}{D_{\max}^2}, \,\, t\in\textstyle\left[t_0,t_0+\frac{D_{\max}}{v}\right].
\end{equation}

After detecting the physical signals generated by the monitored process, sensor $i$ takes the $C_i\sim f_C$ seconds to perform computations and produce its status update. According to Sec.~\ref{sec:system-model}, $f_C$ is a uniform \gls{pdf}; such a model is sufficient to evaluate the delay induced by the computation.

To analyze the delay induced by communication, we evaluate the \gls{pmf} of the time needed for \glspl{sr} and packet transmission, assuming no packet loss. With the scheduling procedure described in Sec.~\ref{sec:system-model}, the probability of the \gls{bs} failing in detecting the \gls{sr} of sensor $i$ is
\begin{equation}
    \zeta_i = 1 - (1+\gamma_i)^{-\frac{1}{\gamma_i}}.
\end{equation}
Due to the independent block-fading channel assumption, $\zeta_i$ is a sufficient statistic to model $m_i$, which follows a geometric distribution, \emph{i.e.}, $m_i \sim \text{Geo}(1 - \zeta_i)$. Therefore, the \glspl{sr} induce a delay of $T_i^\text{SR}=m_iT_\text{f}$, distributed as
\begin{equation}
    T_i^\text{SR} \sim \Pr\{T_i^\text{SR}=mT_\text{f} | 1\leq m\leq M_{\max}\} = \frac{\zeta_i^{m-1}(1-\zeta_i)}{1-\zeta_i^{M_{\max}}},
\end{equation}
for $m\in\{1,\dots,M_{\max}\}$.
A similar analysis can be applied to the packet transmission. From eq.~\eqref{eq:outage-probability}, we have $n_i \sim \text{Geo}(1 - \epsilon_i)$.
Hence, the delay induced by the packet transmission is $T_i^\text{PT}=n_iT_\text{f}$, such that, for $n\in\{1,\dots,N_{\max}\}$,
\begin{align}
    \label{eq:distribution-packet-transmission-attempts}
    T_i^\text{PT} \sim \Pr\{T_i^\text{PT}=nT_\text{f} | 1\leq n\leq N_{\max}\} = \frac{\epsilon_i^{n-1}(1-\epsilon_i)}{1-\epsilon_i^{N_{\max}}}.
\end{align}

The three timing components can be combined to obtain the packet arrival time. Taking into account that sensors acquire frame synchronization through \gls{ss} and considering that the first frame starts at time $t=0$, the packet arrival time is
\begin{equation}
    \label{eq:packet-arrival-time}
    t_i^\text{PA} = \left\lceil\frac{t_i^\text{EA} + C_i}{T_\text{f}}\right\rceil T_\text{f} + T_i^\text{SR} + T_i^\text{PT},
\end{equation}
where the ceiling function ensures the time is aligned with the start of the first available frame after the computation process. We remark that $t_i^\text{PA} = \infty$ if a packet drop event occurs for sensor $i$. Formally, this event is $\mathcal{L}_i=\{m_i > M_{\max} \cup n_i > N_{\max}\}$, having a \emph{packet drop rate} given by the probability 
\begin{equation}
    \label{eq:packet-drop-rate}
    \rho_i = \Pr\{\mathcal{L}_i\} = \zeta_i^{M_{\max}} + \epsilon_i^{N_{\max}} - \zeta_i^{M_{\max}}\epsilon_i^{N_{\max}}.
\end{equation}

According to Definition~\ref{def:pdv}, the \gls{pdv} is obtained from the difference between the arrival times of the last and the first packet received by the \gls{bs}. Therefore, the \gls{pdv} is
\begin{equation}
    \label{eq:packet-delay-variation}
    \Delta_{(I)} = \max_{i\in\mathcal{I}'}\{t_i^\text{PA}\} - \min_{i\in\mathcal{I}'}\{t_i^\text{PA}\},
\end{equation}
where $\mathcal{I}'\subseteq \{1, \dots, I\}$ is the subset of sensors not experiencing a packet drop event.
By definition, $\Delta_{(I)}$ is the \emph{sample range} of $t_i^\text{PA}$; deriving the \gls{pdf} of $t_i^\text{PA}$ allows the use of order statistics tools to determine the distribution of the \gls{pdv} for an arbitrary population of $I$ sensors, leveraging known closed-form expressions for the \gls{pdf} and the \gls{cdf} of the sample range~\cite{Arnold2008}.
In this paper, we aim to provide a simple analytical description of system performance, allowing us to interpret the fundamental behavior of the \gls{twi} framework. To achieve this, we focus on a system with $I = 2$, reducing the complexity of the \gls{pdv} analysis.

\subsection{PDV analysis with two sensors}
Imposing $I = 2$, the \gls{pdv} of eq.~\eqref{eq:packet-delay-variation} can be rewritten as
\begin{equation}
    \label{eq:packet-delay-variation-2-sensors}
    \Delta_{(2)} = |t_1^\text{PA} - t_2^\text{PA}|.
\end{equation}
Deriving a closed-form expression for $\Delta_{(2)}$ remains challenging due to the mix of continuous and discrete random variables and the ceiling function at eq.~\eqref{eq:packet-arrival-time}. To facilitate a tractable design of the \gls{twi}, we propose two approximations for the statistical distribution of $\Delta_{(2)}$: one that applies when the computation delay primarily governs the \gls{pdv} and another suited to cases where event action time is the dominant factor.

\emph{Approximation 1: The computation delay governs the \gls{pdv}.}
This happens when $C_i$ can assume values significantly higher than $t_i^\text{EA}$ and $T_i^\text{SR} + T_i^\text{PT}$.
Thus, the \gls{pdv} is approximated by
\begin{equation}
    \label{eq:approximation-computation-time}
    \Delta_{(2)} \approx \Delta_\text{comp} = |C_1 - C_2|,
\end{equation}
whose \gls{pdf} and \gls{cdf} are given in Lemma~\ref{lem:pdf-approximation-computation-time}.

\begin{lemma}
    \label{lem:pdf-approximation-computation-time}
    The \gls{pdf} and \gls{cdf} of $\Delta_\text{comp}$ are
    \begin{align}
        \label{eq:pdf-approximation-computation-time}
        \textstyle
        f_{\Delta_\text{comp}}(t) &= \frac{2(C_{\max}-C_{\min}-t)}{(C_{\max}-C_{\min})^2}, \,\, t\in[0,C_{\max}-C_{\min}], \\
        \label{eq:cdf-approximation-computation-time}
        F_{\Delta_\text{comp}}(t) &= \begin{cases}
            \frac{2(C_{\max}-C_{\min})t-t^2}{(C_{\max}-C_{\min})^2}, & \hspace{1.3mm}t\in[0,C_{\max}-C_{\min}],\\
            1, & \hspace{1.3mm}t>C_{\max}-C_{\min}.
        \end{cases}
    \end{align}
\end{lemma}
\begin{IEEEproof}
    See Appendix.
\end{IEEEproof}

\emph{Approximation 2: The propagation delay governs the \gls{pdv}, i.e., $t_i^\mathrm{EA} \gg C_i, T_i^\mathrm{SR} + T_i^\mathrm{PT}$.}
This might be the case for big $D_{\max}$ or small $v$. Accordingly, the \gls{pdv} can be approximated by
\begin{equation}
    \label{eq:approximation-propagation-time}
    \Delta_{(2)} \approx \Delta_\text{prop} = |t_1^\text{EA}-t_2^\text{EA}| = \left|\frac{D_1}{v}-\frac{D_2}{v}\right|,
\end{equation}
whose \gls{pdf} and \gls{cdf} are given by Lemma~\ref{lem:pdf-approximation-propagation-time}.

\begin{lemma}
    \label{lem:pdf-approximation-propagation-time}
    The \gls{pdf} and \gls{cdf} of $\Delta_\text{prop}$ are
    \begin{align}
        \label{eq:pdf-approximation-propagation-time}
        \textstyle
        f_{\Delta_\text{prop}}(t) &= \frac{2v^4}{3D_{\max}^4}\hspace{-1mm}\left(2t^3  \hspace{-1mm}- \hspace{-1mm} \frac{6D_{\max}^2}{v^2}t  \hspace{-1mm}+ \hspace{-1mm} \frac{4D_{\max}^3}{v^3}\hspace{-1mm}\right), \,\, t\in\left[0,\textstyle\frac{D_{\max}}{v}\right], \\
        \label{eq:cdf-approximation-propagation-time}
        F_{\Delta_\text{prop}}(t)  \hspace{-1mm}&= \hspace{-1mm} \begin{cases}
            \frac{2v^4}{3D_{\max}^4} \hspace{-1mm} \left(\frac{1}{2}t^4  \hspace{-1mm} - \hspace{-1mm} \frac{3D_{\max}^2}{v^2}t^2  \hspace{-1mm} +  \hspace{-1mm} \frac{4D_{\max}^3}{v^3}t\right), & t\in\big[0,\frac{D_{\max}}{v}\big],\\
            1, & t >\frac{D_{\max}}{v}.
        \end{cases}
    \end{align}            
\end{lemma}
\begin{IEEEproof}
    See Appendix.
\end{IEEEproof}

\begin{figure}[t]
    \centering
    \subfloat[$v=3\cdot10^8$~m/s, $\mathcal{C}=(10,500)$.]{%
        \includegraphics[width=0.5\linewidth]{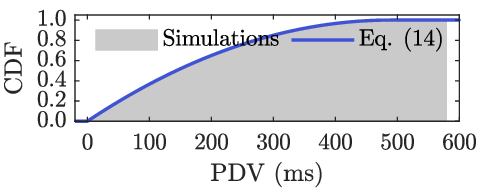}%
        \label{fig:pdv-approximation-computation-time}%
    }%
    \subfloat[$v=300$~m/s, $\mathcal{C}=(10,100)$.]{%
        \includegraphics[width=0.5\linewidth]{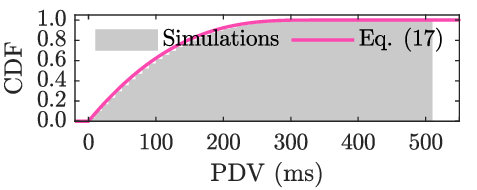}%
        \label{fig:pdv-approximation-propagation-time}%
    }
    \caption{Approximations for the \gls{cdf} of the \gls{pdv} compared with Monte Carlo simulations following eq.~\eqref{eq:packet-delay-variation-2-sensors}. $D_{\max}=100$~m, $T_\text{f}=10$~ms, $M_{\max}=N_{\max}=5$, $\gamma_2=\gamma_\text{TH}=1$, $\mathcal{C}=(C_{\min},C_{\max})$ in ms.}
    \label{fig:pdv-approximation}
\end{figure}

By definition, the \glspl{cdf}~\eqref{eq:cdf-approximation-computation-time} and~\eqref{eq:cdf-approximation-propagation-time} represent the probability that the \gls{pdv} is below a given time, which is essential for verifying if status updates reach the \gls{bs} in time for concurrent delivery toward the application. Based on these and the application requirements, we can determine the optimal \gls{twi} for cases where the approximations hold. To show the approximation accuracy, Fig.~\ref{fig:pdv-approximation} depicts the \glspl{cdf} of Lemmas~\ref{lem:pdf-approximation-computation-time} and~\ref{lem:pdf-approximation-propagation-time} along with \gls{pdv}'s empirical distributions, obtained through Monte Carlo simulations using eq.~\eqref{eq:packet-delay-variation-2-sensors}.

\paragraph*{TWI design} With $I=2$, the simultaneity violation happens if, upon the reception of one status update packet, the packet of the other is dropped by its sensor or arrives at the \gls{bs} after the \gls{twi} duration.\footnote{No simultaneity violation occurs if both sensors drop their packets, as no updates will be received by the \gls{bs} to trigger the start of a \gls{twi}.} Without loss of generality, assuming that the \gls{bs} received first from sensor~1, the \gls{psv} is
\begin{equation}
    \label{eq:psv}
    \sigma(W) = \Pr\{\mathcal{L}_2 \cup \Delta_{(2)}>W\} = 1 - (1-\rho_2)\Pr\{\Delta_{(2)}<W\},
\end{equation}
where $\mathcal{L}_2$ and $\rho_2$ are the packet drop event and rate for sensor $i=2$---see eq.~\eqref{eq:packet-drop-rate}. Remark that $\sigma \rightarrow \rho_2$ when $\Pr\{\Delta_{(2)}<W\} \rightarrow 1$, meaning that $\rho_2$ is a lower bound of the \gls{psv}.
By replacing $\Pr\{\Delta_{(2)}<W\}$ in eq.~\eqref{eq:psv} with \eqref{eq:cdf-approximation-computation-time}, we get the \gls{psv} when the computation delay governs the \gls{pdv}, that is
\begin{equation}
    \label{eq:psv-computation-approximation}
    \sigma_\text{comp}(W) = 1 - (1-\rho_2)F_{\Delta_\text{comp}}(W).
\end{equation}
Conversely, by replacing it with eq.~\eqref{eq:cdf-approximation-propagation-time}, we get the \gls{psv} when the propagation delay governs the \gls{pdv}, that is
\begin{equation}
    \label{eq:psv-propagation-approximation}
    \sigma_\text{prop}(W) = 1 - (1-\rho_2)F_{\Delta_\text{prop}}(W).
\end{equation}
By inverting eqs.~\eqref{eq:psv-computation-approximation} and~\eqref{eq:psv-propagation-approximation}, we can design the \gls{twi} duration satisfying a target \gls{psv}.

%
%
\section{Numerical Results}\label{sec:numerical-results}

In this section, we deploy a performance analysis of the system through numerical simulations. We assume the physical process event occurs at $t_0=0$ for all results. Moreover, realizations of the random variables are computed by sampling the random distributions defined in Sec.~\ref{sec:twi-design}.

\begin{figure}[t]
    \centering
    \subfloat[$v=3\cdot10^8$~m/s, $\mathcal{C}=(10,500)$.]{%
        \includegraphics[width=0.48\linewidth]{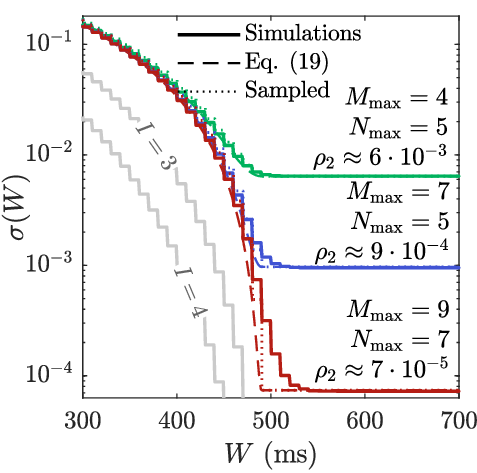}%
        \label{fig:psv-approximation-computation-time}%
    }%
    \hfill%
    \subfloat[$v=300$~m/s, $\mathcal{C}=(0,10)$.]{%
        \includegraphics[width=0.48\linewidth]{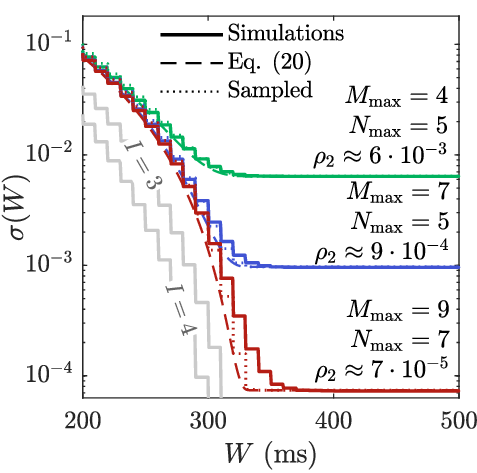}%
        \label{fig:psv-approximation-propagation-time}%
    }%
    \caption{\gls{psv} calculated by the approximations of (a)~eq.~\eqref{eq:psv-computation-approximation} and (b)~eq.~\eqref{eq:psv-propagation-approximation}, and obtained from Monte Carlo simulations. $D_{\max}=100$~m, $T_\text{f}=10$~ms, $\gamma_2=4$, $\gamma_\text{TH}=1$, $\mathcal{C}=(C_{\min},C_{\max})$ in ms.}
    \label{fig:psv}
\end{figure}

The comparison between the analytical and empirical \gls{psv} is depicted in Fig.~\ref{fig:psv}, tested under different communication parameters.
Specifically, the \gls{psv} obtained in the Monte Carlo simulations is compared with the approximation of eq.~\eqref{eq:psv-computation-approximation} in Fig.~\ref{fig:psv-approximation-computation-time} and~\eqref{eq:psv-propagation-approximation} in Fig.~\ref{fig:psv-approximation-propagation-time}. Moreover, we also present the approximated \gls{psv} sampled at multiples of the frame duration---\emph{i.e.}, $\sigma_\mathrm{comp}(k T_\mathrm{f})$ and $\sigma_\mathrm{prop}(k T_\mathrm{f})$, $k\in\mathbb{N}$---due to the fact that the frame-based communication process limits $W$ to the frame boundaries.
The curves show the effect of $\rho_2$ bounding the \gls{psv}, meaning that a full system optimization must account not only for the \gls{twi} duration but also for the communication parameters. 
For $\rho_2 \ge 10^{-4}$, the approximations accurately model the \gls{psv} for all values of $W$, enabling the design of $W$ based on a target $\sigma$. However, for $\rho_2 < 10^{-4}$, the approximation loses accuracy around the values of $W$ where $\sigma_\mathrm{comp}$ and $\sigma_\mathrm{prop}$ approach the bound---precisely, for $W = C_{\max} - C_{\min}$ in eq.~\eqref{eq:cdf-approximation-computation-time} (Fig.~\ref{fig:psv-approximation-computation-time}), and $W = D_{\max}/v$ in eq.~\eqref{eq:cdf-approximation-propagation-time} (Fig~\ref{fig:psv-approximation-propagation-time}). This discrepancy arises because the impact of communication delays for high values of $M_\mathrm{max}$ and $N_\mathrm{max}$ slows the convergence of $\Pr\{\Delta_{(2)} < W\}$ toward $1$---see eq.~\eqref{eq:psv}.\footnote{Future work will focus on deriving a tighter upper bound of the \gls{psv} for robust \gls{twi} design in ultra-reliable scenarios.}

Fig.~\ref{fig:psv} also shows the \gls{psv} for $I>2$, conditioned on successful reception. For fairness, only the first two received packets are considered. Eqs.~\eqref{eq:psv-computation-approximation} and~\eqref{eq:psv-propagation-approximation} bound the \gls{psv} for such cases, applying to worst-case analysis. Meanwhile, tighter approximations can be derived using order statistics.

\begin{figure}[t]
    \centering
    \subfloat[$D_{\max}=100$ m, $v=3\cdot10^8$ m/s, $\mathcal{C}_k=(C_{\min},C_{\max})$ in ms.]{%
        \includegraphics[width=\linewidth]{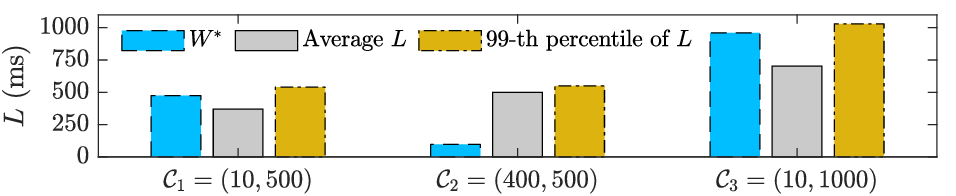}%
        \label{fig:latency-approximation-computation-time}%
    }
    
    \subfloat[$C_{\min}=0$, $C_{\max}=10$ ms, $\mathcal{P}_k=(D_{\max},v)$ in m and m/s.]{%
        \includegraphics[width=\linewidth]{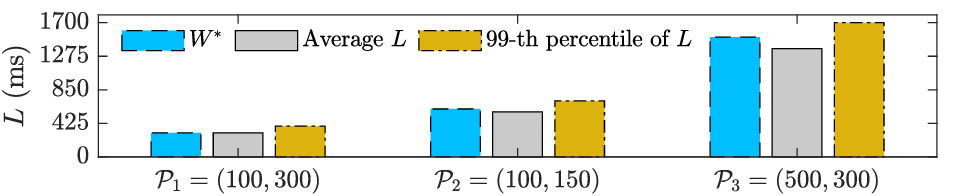}%
        \label{fig:latency-approximation-propagation-time}%
    }
    \caption{\gls{twi} computed by the approximations of (a)~eq.~\eqref{eq:psv-computation-approximation} and (b)~eq.~\eqref{eq:psv-propagation-approximation} to achieve $\sigma(W^*)=10^{-3}$ and resulting latency obtained in different setups. $T_\text{f}=10$~ms, $\gamma_2=4$, $\gamma_\text{TH}=1$, $M_{\max}=9$, $N_{\max}=7$, $\rho_2=7\cdot10^{-5}$.}
    \label{fig:latency}
\end{figure}

Using the statistical model from Sec.~\ref{sec:twi-design}, we show how the delivery of status updates affects the application. Here, \emph{latency} refers to the time from $t_0$ to the delivery of the status updates to the application within the first available \gls{twi}. We analyze two cases: 1) when the packet of sensor~2 arrives before the end of the \gls{twi}, the updates are concurrently delivered at the time $t_1^\text{PA} + \Delta_{(2)}$; and, 2) when the sensor~2's packet is received in a different \gls{twi}, the first update is delivered alone at the time $t_1^\text{PA} + W$. Thus, the latency is given by $L = t_1^\text{PA} + \min\{\Delta_{(2)},W\} - t_0$, which depends not only on the \gls{twi} duration, but also on the random delays captured by $\Delta_{(2)}$.
Fig.~\ref{fig:latency} shows the values of $L$ obtained with different system parameters. By evaluating multiple setups, we can verify how the physical process and computation parameters affect the definition of $W$ and the induced $L$. Notably, the average $L$ can be lower than $W$, as the random delays can make the second update arrive before the end of the \gls{twi}.

%
%
\section{Discussion and Conclusions}\label{sec:conclusions}

In this paper, we have addressed the problem of event chronology in the reception of status updates of sensors observing the same physical process. By modeling the random delays involved in the generation and transmission of updates, we proposed a \gls{twi} duration design to attain a target \gls{psv} in a two-sensor scenario, validating it through numerical simulations. 
Our framework is well-suited for evaluating closed-loop latency in control tasks that rely on chronologically ordered, distributed observations. 
Future works will extend the analysis to multisensor scenarios, where the assumption of collision-free access poses a scalability challenge: generating orthogonal sequences for massive sensor populations becomes infeasible with a bounded frame duration. Addressing collision-induced communication delays is a crucial direction for future research.

%
%
\appendix

%
%

To develop the proof of Lemma~\ref{lem:pdf-approximation-computation-time} and~\ref{lem:pdf-approximation-propagation-time}, we use the \gls{cdf} method to obtain the \gls{pdf} of the absolute difference between two random variables. Let us define the random variables $X_1, X_2\sim f_X$ and $Y=|X_1-X_2|\sim f_Y$.
Thus, we have
\begin{align}
    \nonumber
    F_Y(y) &= \Pr\{|X_1-X_2|<y\} = \Pr\{-y<X_1-X_2<y\},\\
    \label{eq:cdf-Y}
    &= \textstyle\int_{-\infty}^\infty\int_{x_2-y}^{x_2+y} f_X(x_1)~f_X(x_2)~dx_1~dx_2.
\end{align}
The \gls{pdf} of $Y$ is obtained derivating~\eqref{eq:cdf-Y} w.r.t. $y$, obtaining
\begin{equation}    
    \label{eq:pdf-Y}
    f_Y(y) = \textstyle\int_{-\infty}^\infty [f_X(x_2+y)+f_X(x_2-y)]~f_X(x_2)~dx_2.
\end{equation}

In Lemma~\ref{lem:pdf-approximation-computation-time}, we have $X_1=C_1$, $X_2=C_2$, $Y=\Delta_\text{comp}$, and $f_X=\mathcal{U}(C_{\min}, C_{\max})$.
By substituting these quantities and solving the integral, we obtain $f_{\Delta_\text{comp}}$ as in~\eqref{eq:pdf-approximation-computation-time}. Then, by integrating eq.~\eqref{eq:pdf-approximation-computation-time} w.r.t. $t$, we obtain $F_{\Delta_\text{comp}}$ as in~\eqref{eq:cdf-approximation-computation-time}.

%
%

In Lemma~\ref{lem:pdf-approximation-propagation-time}, $X_1=D_1/v$, $X_2=D_2/v$, $Y=\Delta_\text{prop}$, and $f_X=f_D$ as in~\eqref{eq:pdf-D}.
By substituting these quantities  in the integral, we obtain $f_{\Delta_\text{prop}}$ as in~\eqref{eq:pdf-approximation-propagation-time}, while $F_{\Delta_\text{prop}}$ as in~\eqref{eq:cdf-approximation-propagation-time}, is obtained by integrating eq.~\eqref{eq:pdf-approximation-propagation-time} w.r.t. $t$, completing the proof.

%
%

\end{document}